\documentclass[twocolumn,showpacs,preprintnumbers,amsmath,amssymb,aps,a4paper]{revtex4}

\usepackage{graphicx}
\usepackage{dcolumn}
\usepackage{bm}
\usepackage{color}

\newcommand{\ket}[1]{\left|{#1}\right\rangle}

\begin{document}

\title{Stimulated Raman Adiabatic Passage in Tm$^{3+}$:YAG}

\author{A. L. Alexander}
\email{annabel.alexander@gmail.com}
\author{R. Lauro}
\author{A. Louchet}
\author{T. Chaneli\`{e}re}
\author{J. L. Le Gou\"{e}t}

 \affiliation{Laboratoire Aim\'{e} Cotton, CNRS-UPR 3321, Univ Paris-Sud, B\^{a}t. 505, 91405 Orsay cedex, France.}

\begin{abstract}We report on the experimental demonstration of stimulated Raman adiabatic passage (STIRAP) in a Tm$^{3+}$:YAG crystal.  Tm$^{3+}$:YAG is a promising material for use in quantum information processing applications, but as yet there are few experimental investigations of coherent Raman processes in this material.  We investigate the effect of inhomogeneous broadening and Rabi frequency on the transfer efficiency and the width of the two-photon spectrum.  Simulations of the complete Tm$^{3+}$:YAG system are presented along with the corresponding experimental results.
\end{abstract}


\maketitle

\section{Introduction}

Rare earth ions doped into inorganic crystals have recently been considered for use in the field of quantum information processing \cite{Longdell2004a,Ohlsson2002,Nilsson2005,Longdell2005,Ichimura2001,Hetet2008}. These materials are promising as they offer very long coherence times in both the hyperfine \cite{Fraval2005} and optical transitions \cite{Equall1994}. Currently the majority of quantum information processing demonstrations have been performed in crystals doped with either praseodymium or europium. Unfortunately these materials can only be optically addressed with dye-lasers.  The task to stabilize these lasers to the sub-kHz linewidths required for quantum information processing experiments is so challenging that only a few groups in the world have achieved such laser sources \cite{Klieber2003, Sellars1994,Julsgaard2007}.  

Another disadvantage of praseodymium and europium doped crystals is the small hyperfine splitting, which is of the order of ten's of MHz.  This small separation limits possible quantum information storage demonstrations in these crystals as the bandwidth of existing quantum sources does not match this narrow hyperfine splitting.  It is therefore worthwile to investigate other rare earth ions which do not suffer these disadvantages.

One such promising rare earth ion is thulium.  The wavelength of thulium is 793~nm which falls in the range of easily stabilized semi-conductor lasers.   The only isotope of thulium has an $I=1/2$ nuclear spin but unfortunately does not exhibit hyperfine structure due to $J$ quenching \cite{Abragam1970}.  It has proven possible to create a 3-level $\Lambda$-system in this material via the application of a magnetic field with a particular orientation \cite{Louchet2007}. The advantage of this $\Lambda$-system is that it possesses an adjustable ground state hyperfine splitting via the magnitude of the applied magnetic field.  Thus Tm$^{3+}$:YAG does not suffer from a limited bandwidth for possible quantum information applications and is optically accessible with easily stabilized lasers.  

There has been a startling lack of investigation into Raman coherent processes in this material.  The first nuclear spin coherence via optical excitation has only recently been demonstrated \cite{Louchet2008} and it is certainly obvious that these types of processes need to be studied further if this material is to be useful for quantum information processing applications.  Stimulated Raman adiabatic passage (STIRAP) \cite{Gaubatz1990,Schiemann1993} is one such coherent process which deserves attention in this material.  It provides complete coherent population transfer in a $\Lambda$-system without suffering from radiative losses and is robust against slight pulse variations.  STIRAP is thus important in the field of quantum information processing for use in the preparation and manipulation of ions.  

In this paper we will first discuss the theory of STIRAP using the simple 3-level $\Lambda$-system at two-photon resonance.  We will then present numerical simulations of the full Tm:$^{3+}$:YAG level structure alongside simulations of the simple 3-level $\Lambda$-system.  Following this will be a presentation of experimental and numerical results along with discussion and conclusions.

\section{Theory}
\begin{figure}[!ht]
\begin{centering}
\includegraphics[width=3cm]{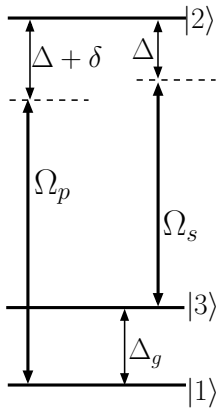}
\caption{Three level $\Lambda$-system where the ground state hyperfine levels are split by $\Delta_{g}$.  The optical, or one-photon, detuning is labelled $\Delta$ and is the frequency difference between the Stokes pulse and the transition $\ket{3}$-$\ket{2}$.  The two-photon detuning is labelled $\delta$ and is the frequency difference between the pump pulse and two-photon resonance.  The Rabi frequencies of the pump and Stokes pulses are labelled $\Omega_{p}$ and $\Omega_{s}$ respectively.}\label{Lambda}
\end{centering}
\end{figure}

For a simplified discussion of STIRAP we consider the 3-level $\Lambda$-system shown in Fig.~\ref{Lambda}.  The objective is to efficiently transfer population from level $\ket{1}$ to level $\ket{3}$ via the use of two temporally overlapping laser pulses.   

The ground state hyperfine levels are split by $\Delta_{g}$.  The detuning between the Stokes pulse and optical resonance is denoted $\Delta$ and is referred to as the optical or one-photon detuning.  The detuning of the pump pulse from two-photon resonance is denoted $\delta$ and corresponds to the two-photon detuning.  In the simplified case we assume that $\delta=0$, ie. we are at two-photon resonance.  The system is initially prepared with all the population in level $\ket{1}$ and two, non-simultaneous, laser pulses are applied.  The Stokes pulse, applied first, excites the, initially empty, transition $\ket{3}-\ket{2}$ and the pump pulse, appplied temporally overlapping the Stokes pulse, excites the transition $\ket{1}-\ket{2}$.  The pump and Stokes pulses have the following, respective, Rabi frequencies: $\Omega_{p}=-\mu_{12}\varepsilon_{p}(t)/\hbar$ and $\Omega_{s}=-\mu_{32}\varepsilon_{s}(t)/\hbar$.  The dipole transition moments are the $\mu_{ij}$'s and the applied, time-dependent, electric fields of the laser pulses are the $\varepsilon_{s,p}(t)$'s.  

The eigenstates of the instantaneous RWA Hamiltonian of this system are well known and are given by the following linear combinations of the bare states $\ket{1}$, $\ket{2}$ and $\ket{3}$:
\begin{eqnarray}
\ket{a^{+}}&=&\sin\theta\sin\phi\ket{1}+\cos\phi\ket{2}+\cos\theta\sin\phi\ket{3} \nonumber \\
\ket{a^{0}}&=&\cos\theta\ket{1}-\sin\theta\ket{3} \nonumber \\
\ket{a^{-}}&=&\sin\theta\cos\phi\ket{1}-\sin\phi\ket{2}+\cos\theta\cos\phi\ket{3}
\end{eqnarray}
where the time-dependent mixing angles, $\theta$ and $\phi$, are given by:
\begin{eqnarray}
\tan\theta&=&\frac{\Omega_{p}(t)}{\Omega_{s}(t)} \nonumber \\
\tan2\phi&=&\frac{\sqrt{\Omega_{p}^{2}(t)+\Omega_{s}^{2}(t)}}{\Delta}
\end{eqnarray}
and we have the following time-dependent, dressed state eigenvalues:
\begin{eqnarray}
\omega^{0}&=&0 \nonumber \\
\omega^{\pm}(t)&=&\Delta/2\pm 1/2\sqrt{\Delta^{2}+\Omega_{p}^{2}(t)+\Omega_{s}^{2}(t)}
\end{eqnarray}

After examining the eigenstates it is obvious that two dressed states, $\ket{a^{\pm}}$, contain the `leaky' bare state $\ket{2}$.  Thus these dressed states cannot be used for complete population transferral as some population will be promoted to the excited level, $\ket{2}$, where it will experience radiative losses and therefore be lost to the final state $\ket{3}$.  

Inspection of the dressed state $\ket{a^{0}}$, known as the `dark' state as it contains no component of the electronic excited state, demonstrates that it should be possible to perform efficient population transferral using this state.  If the Stokes pulse is applied before the pump pulse then at time $t\rightarrow -\infty$ we have the mixing angle $\theta=0^{\circ}$ and the dressed state $\ket{a^{0}}$ corresponds to the bare, populated, state $\ket{1}$.  At the end of the interaction, at time $t\rightarrow\infty$, where the pump pulse has been applied after the Stokes pulse, we have a mixing angle of $\theta=90^{\circ}$ and the state $\ket{a^{0}}$ corresponds to the bare state $\ket{3}$.  

This implies that provided the evolution of the system is adiabatic and that the Stokes pulse preceedes the pump pulse the system will remain in the $\ket{a^{0}}$ state, or the `dark' state, throughout the interaction. Hence the population initially in level $\ket{1}$ will be transferred into level $\ket{3}$ whilst never populating level $\ket{2}$.  Thus the population transferral does not depend on either the optical coherence lifetime or the decay rate from the upper level provided the adiabacity condition is satisfied. 

The adiabaticity condition requires the energy spacing between the eigenvalues to be much larger than the dynamic coupling term, given by $\dot{\theta}(t)$, where the overdot represents the time derivative.  The adiabacity condition is \cite{Kuklinski1989}:
\begin{equation}
\left|\frac{\dot{\Omega}_{p}\Omega_{s}-\Omega_{p}\dot{\Omega}_{s}}{\Omega_{p}^{2}+\Omega_{s}^{2}}\right|\ll\left|\frac{\Delta}{2}\pm\frac{\sqrt{\Delta^{2}+\Omega_{p}^{2}(t)+\Omega_{s}^{2}(t)}}{2}\right| \label{adiabat} 
\end{equation}
Thus for any given shape of the laser pulses, $\Omega_{p}$ and $\Omega_{s}$, the adiabacity condition can be evaluated at any time, $t$. 

At optical resonance this adiabacity condition is generally satisfied, for smooth pulses, if:
\begin{equation}
\Omega_{eff}\tau\gg 1 
\label{near_cond}
\end{equation} 
where $\tau$ is the pulse duration and the effective Rabi frequency is: 
\begin{equation}
\Omega_{eff}=\sqrt{\Omega_{p}^{2}+\Omega_{s}^{2}}
\end{equation}
Far from optical resonance the adiabatic condition, for smooth pulses, becomes instead:
\begin{equation}
\Omega_{eff}^{2}\tau\gg \left|\Delta\right|
\label{far_cond}
\end{equation}
which sets both an upper limit on the optical detuning and a lower limit on the Rabi frequencies for which population transferral is efficient.  Therefore carefully selected pump and Stokes pulses, tuned to two-photon resonance, will transfer population adiabatically over a spectral range given by the adiabacity condition of Eqn.~\ref{far_cond}.

The eigenvectors and eigenvalues of the Hamiltonian involving a non-zero two-photon detuning, $\delta$, are much more complicated than those of the two-photon resonance case.  These eigenvectors and eigenvalues are explicitly solved in Ref.~\cite{Fewell1997} in which it is demonstrated that all three eigenvectors contain a contribution of the excited state level.  It was also shown that away from two-photon resonance there is no adiabatic state which connects the initial bare state at the beginning of the sequence with the desired final bare state at the end of the interaction \cite{Fewell1997}.  This implies that adiabatic processes are not the means for population transferral in this type of system.  When two-photon detuning is present the majority of the population is instead transferred via diabatic processes which involve transisitions between the adiabatic states.  

\section{Simulations}

Simulations of the Tm$^{3+}$:YAG system were performed in order to aid our understanding of the system.  The full Tm$^{3+}$:YAG level scheme is actually quite simple when compared to that of the full praseodymium and europium systems.  Praseodymium and europium both have a nuclear spin of $I=5/2$ and thus when in a magnetic field the level scheme consists of six energy levels in both the ground and excited electronic states.  Thulium on the other hand has a nuclear spin of $I=1/2$ and therefore when it is placed in a magnetic field the energy level structure is much simpler, consisting of only two Zeeman energy levels in both the ground and excited electronic states.  

\begin{figure}[!ht]
\begin{centering}
\includegraphics[width=5cm]{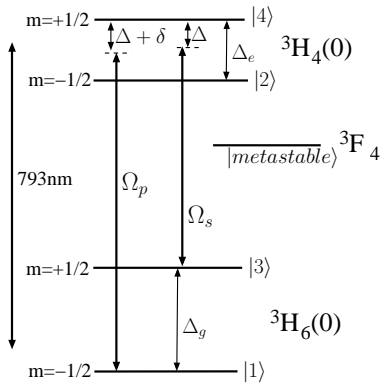}
\caption{Level structure of Tm$^{3+}$:YAG in a magnetic field, the ground and excited electronic states are split into two Zeeman levels which are labelled with their nominal nuclear spin numbers.  The ground and excited state hyperfine splittings, labelled $\Delta_{g}$ and $\Delta_{e}$ respectively, are dependent on the magnitude of the applied magnetic field, with $\Delta_{e}=\Delta_{g}/2.5$.  The two-photon detuning and the Rabi frequencies of the transitions are as described in the 3-level system.  The optical detuning is the frequency difference between the Stokes pulse and the transition $\ket{3}$-$\ket{4}$.}\label{4level}
\end{centering}
\end{figure}

This simple energy level structure of Tm$^{3+}$:YAG, displayed in Fig.~\ref{4level}, is thus easy to completely model, even when including the additional metastable energy level.  
In the full Tm$^{3+}$:YAG level structure the ground and excited state splittings are labelled as $\Delta_{g}$ and $\Delta_{e}$ respectively, and are dependent on the magnitude of the applied magnetic field, with $\Delta_{e}$ = $\Delta_{g}/$2.5.  The optical detuning, $\Delta$, and the two-photon detuning, $\delta$, are as defined in the previous section.

One complication to the Tm$^{3+}$:YAG model arises from the fact that the Stokes and pump pulses can excite all four optical transitions.  This is possible due to the lack of polarisation selectivity of the optical transitions and the mixing of the nuclear spin states, which relaxes the spin selection rule, $\Delta m=0$.  This relaxation results in the spin-flip transitions, $\ket{3}$-$\ket{2}$ and $\ket{1}$-$\ket{4}$, having a transition dipole moment of 0.37 times that of the transitions not involving a spin-flip \cite{Seze2006}.  The nominal nuclear spin labels in Fig.~\ref{4level} thus represent the strong and weak transitions.

Whilst the Stokes and pump pulses are applied with equal intensities the Rabi frequencies of the transitions are different depending on which transition is being excited.  The Rabi frequency of the weak transition is 0.37 times the value of the Rabi frequency of the strong transition.  In all the following discussions the Rabi frequency listed is that of the strong transition.  
  
The metastable state in Fig.~\ref{4level} is included as a kind of reservoir in the model.  This state has a much longer radiative lifetime, of 10~ms \cite{Macfarlane1993a}, compared to that of the optical excited level, of 800~$\mu$s \cite{Macfarlane1993}.  In addition to this, most of the population in the optical excited level decays to the metastable state rather than the ground state, in a ratio of 3:1 \cite{Seze2005}.

The optical Bloch equations for this full energy level scheme, including the metastable state and the fact that multiple transitions are excited by the pump and Stokes pulses, are numerically solved in the interaction picture.  We have also included population and coherence decay terms in the equations and the transfer efficiency is determined 300~$\mu$s after the application of the pump pulse.  The Stokes and pump pulses both have a Gaussian lineshape with a FWHM (of intensity) of 30/$\sqrt{2}$~$\mu$s and a pulse delay of -17~$\mu$s, with the Stokes pulse preceding the pump pulse. 

\begin{figure}[!ht]
\centering
\includegraphics[width=9cm]{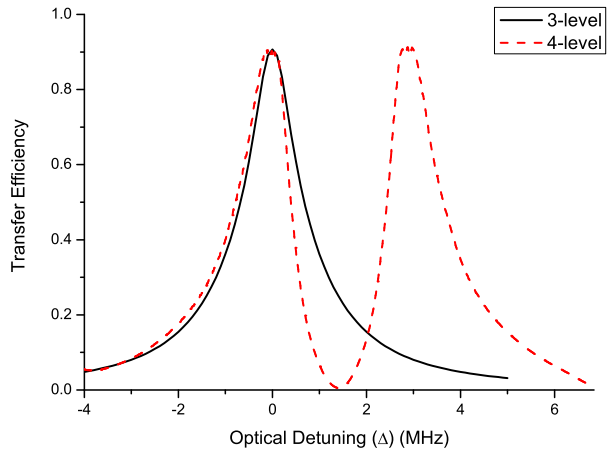}
\caption{(colour online) The optical detuning scan from the 4-level model, dashed line (red), and the standard 3-level model, solid line (black).  In this case the excited state splitting was 2.84~MHz, the Rabi frequency on the strong transition was 510~kHz, the pulse duration was 30/$\sqrt{2}$~$\mu$s and the pulse delay was -17~$\mu$s.  These models were both performed at two-photon resonance and we record the transfer efficiency at each optical detuning value, $\Delta$.}\label{NoInt_transfer_width}
\end{figure}

In Fig.~\ref{NoInt_transfer_width} we compare the optical detuning scans obtained from the 4-level model described above and the standard 3-level model of the level structure shown in Fig.~\ref{Lambda}.  This 3-level model is simply the numerical solution to the standard optical Bloch equations where we have included dephasing and relaxation terms and the fact that we have a strong and weak transition.  Both these simulations were performed at two-photon resonance, ie. at $\delta=0$.  In this case the transfer efficiency is determined from the population difference between the excited level, level $\ket{4}$ in the 4-level model, and level $\ket{3}$.    

Looking at only the one-photon linewidth centred at $\Delta$~=~0MHz in Fig.~\ref{NoInt_transfer_width} for the 3-level and 4-level models we see that the introduction of the fourth level and the multiple excitations of the transitions results in an asymmetry to the one-photon lineshape as well as reducing the overall linewidth.  The one-photon linewidth (FWHM) of the 4-level model is 1.3~MHz as compared to 1.5~MHz for the 3-level model.  From the particular pulse parameters used here the adiabacity criterion of Eqn.~\ref{far_cond} reduces to 6.3~MHz $\gg\left|\Delta\right|$.  The linewidths of both the models agree with this criterion but this equation does not explain the asymmetry to the 4-level model.

In Fig.~\ref{NoInt_transfer_width} we also see the contributions from both $\Lambda$-systems in the 4-level model.  At $\Delta$~=~0~MHz we are on-resonance with the excited state level $\ket{4}$ whilst at $\Delta$~=~2.48~MHz we are now on-resonance with the other $\Lambda$-system, with the excited state level $\ket{2}$.  It can be seen in this figure that the maximum transfer efficiency is the same for both $\Lambda$-systems even though the Rabi frequencies of the Stokes and pump pulses have been switched.  It is thought that the transfer efficiency dramatically decreases at $\Delta$~=~1.24~MHz, the midway point between the two excited state levels, due to an interference effect between these two $\Lambda$-systems.  It is this interference effect which causes the asymmetry and reduced linewidth of the 4-level model over the 3-level.

\begin{figure}[!ht]
\begin{centering}
\includegraphics[width=10cm]{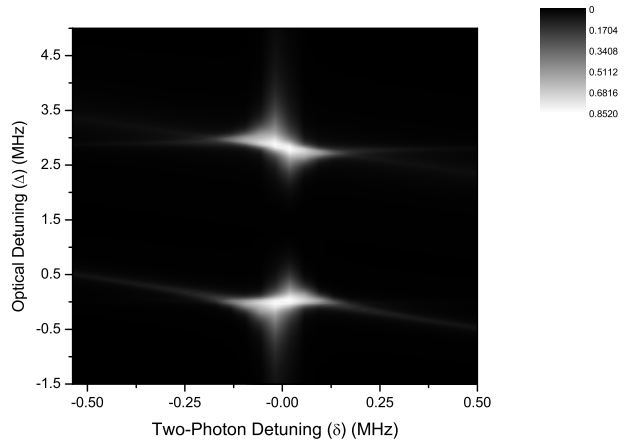}
\caption{The transfer efficiency determined from the 4-level model for different values of the two-photon and optical detunings for an excited state splitting of 2.84~MHz, a Rabi frequency of 300~kHz and a pulse delay of -17~$\mu$s.}\label{contour}
\end{centering}
\end{figure}

In Fig.~\ref{contour} is shown the transfer efficiency determined from the 4-level model as both the one- and two-photon detunings are varied.  The effect of the second excited state hyperfine level can be seen in this figure.  
The efficient population transfer centered at $\Delta=0$ is due to the pump and Stokes pulses being optically resonant with level $\ket{4}$ of Fig.~\ref{4level}.  Whilst the efficient transfer centered at $\Delta=$~2.84~MHz is due to the pump and Stokes pulses being applied nearly resonant with level $\ket{2}$ of Fig.~\ref{4level}, the second excited state hyperfine level.  The lack of population transfer for optical detuning values between these two levels is thought to be due to the interference effect of the two $\Lambda$-systems.

We also compare the standard 3-level model and the 4-level model described above for a two-photon detuning scan.  In this case we perform the simulations at optical resonance, ie. at $\Delta=0$, whilst we vary the frequency difference between the pump pulse and two-photon resonance, $\delta$.  
\begin{figure}[!ht]
\begin{centering}
\includegraphics[width=9cm]{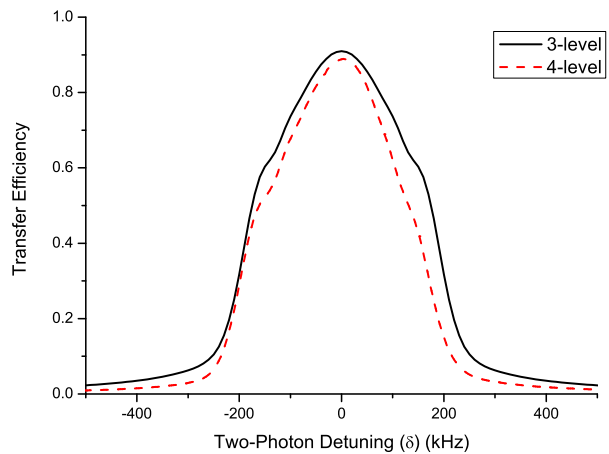}
\caption{(colour online) Two-photon detuning scan at optical resonance, $\Delta=0$.  The solid line (black) is the simulation of the standard 3-level model and the dashed line (red) is the result from the 4-level model.  In these models the Rabi frequency on the strong transition was 400~kHz, the ground state splitting was 7.1~MHz, the excited state splitting was 2.84~MHz and the pulse delay was -17~$\mu$s.}\label{Scan}
\end{centering}
\end{figure}

We see in Fig.~\ref{Scan} that the addition of the fourth level makes less difference to the two-photon detuning scan than it did for the optical detuning scan.  The two-photon linewidth is again slightly narrower for the 4-level model. The two-photon linewidth (FWHM) of the 3-level model is 360~kHz whilst that of the 4-level is 320~kHz and there is again a slight asymmetry to the 4-level lineshape.  Adiabatic population transferral can only occur at two-photon resonance.  Thus, we see in Fig.~\ref{Scan} that there is a large amount of population transferral due to non-adiabatic, or diabatic, transitions in both models.  

\section{Experimental Set-Up}

\begin{figure}[!ht]
\begin{centering}
\includegraphics[width=7cm]{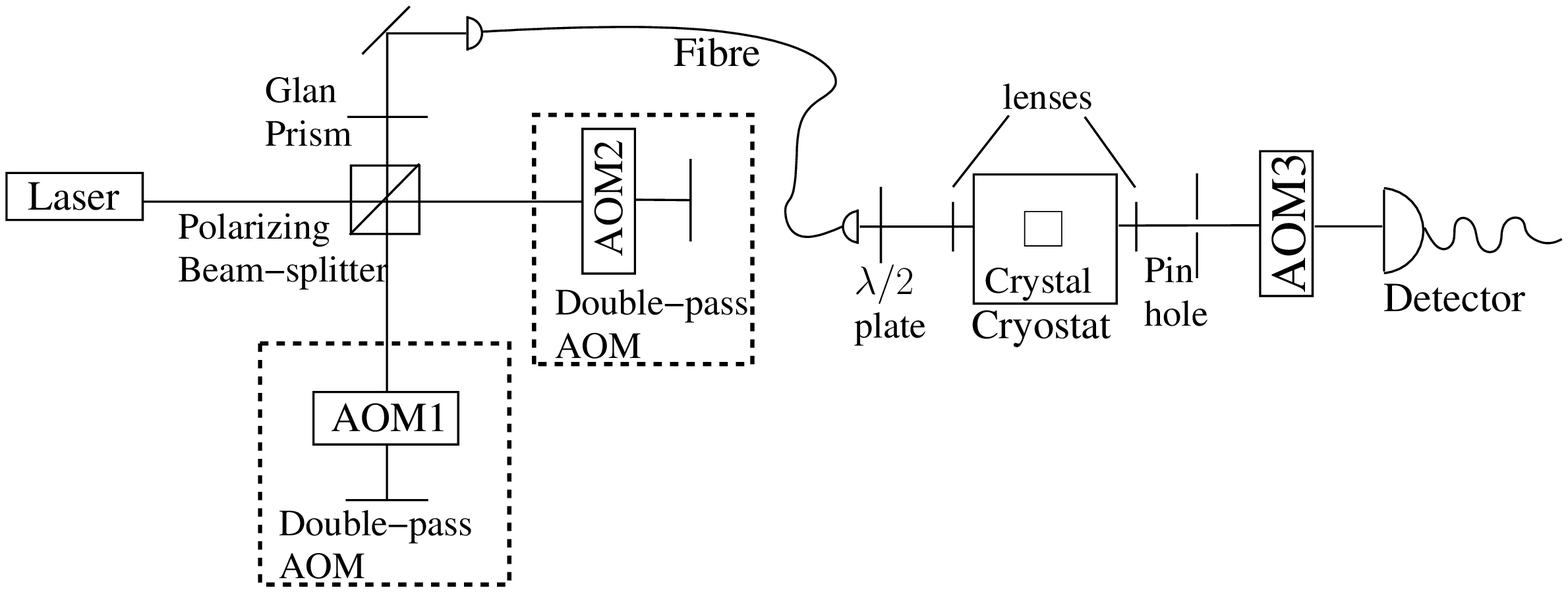}
\caption{Experimental set-up of the system, acousto-optic modulators are labelled as AOM.  AOM1 and 2 are arranged in a double-pass arrangement so the light on the crystal is not spatially shifted when the frequency changes.  These double-pass AOMs are centred on 110~MHz (AA OptoElectronics) and provide the pump, Stokes, probe and preparation pulses.  AOM3 is used as a gate to protect the detector from the intense preparatory and Gaussian-shaped pulses.}\label{exp}
\end{centering}
\end{figure}

When performing STIRAP experimentally the system is excited with an extended cavity semi-conductor laser operating at 793~nm.  It is stabilized with a high-finesse Fabry-Perot cavity via a Pound-Drever-Hall servoloop with a frequency stabilisation of 200~Hz over 10~ms \cite{Crozatier2004}. The laser is then amplified with a semi-conductor tapered amplifier (Toptica BoosTA).  The experimental set-up is shown in Fig~\ref{exp}.  AOMs 1 and 2 are arranged in a double-pass configuration so that the light on the sample is not spatially shifted with a frequency shift.  These AOMs are driven by a dual-channel 1~Gigasample/s arbitrary waveform generator (Tektronix AWG520) which can produce arbitrary amplitude and phase shaping.  

After passing back through the polarizing beam-splitter a common polarization is given to both components via a Glan prism.  This recombined beam is then coupled into a 2~m long single-mode fibre.  The light polarization direction is optimized with a half-wave plate to ensure maximum Rabi frequency.  This beam is then focussed onto the 0.1~at.\% Tm$^{3+}$:YAG sample which is 5~mm in the light propagation direction.  The sample is cooled to 1.8~K in an Oxford 6T Spectromag SM4 cryostat.  

After the cryostat the beam passes through a 50~$\mu$m pinhole and is then imaged onto an avalanche photodiode (Hamamatsu C5460).  The Gaussian shaped cross-section of the laser beam means that those ions positioned at the edge of the beam diameter do not experience the maximum Rabi frequency.  The pinhole ensures only those ions in the centre of the beam and thus subjected to the maximum Rabi frequency of the pump and Stokes pulses are probed.  A final AOM protects the detector from strong light pulses.  

The applied magnetic field, generated by superconducting coils, is applied in the direction optimizing the branching ratio \cite{Louchet2007}.  The magnitude of this applied magnetic field determines the splitting of the hyperfine levels.  The ground state splitting is given by 41~MHz/T with an excited state splitting of 16~MHz/T \cite{Louchet2008}.  The magnitude of this applied field was of the order of 0.2-0.5~T.   

The inhomogeneous width of the optical transition is $\sim$25~GHz which is much larger than the hyperfine level splitting.  Thus a single laser frequency can drive transitions in different ensembles of ions at the same time.  In order to drive and probe the STIRAP process in a single ensemble of thulium ions with a well-defined transition frequency we perform a preparation sequence based on optical hole burning.  

The preparation sequence consists of repeated burning over a range of 3~MHz via scanning the laser.  This optical pumping creates a trench in the absorption profile.  After this preparation sequence the pump and Stokes pulses are applied.  The frequency of the Stokes pulse is chosen such as to place it in the centre of this created transmission window and thus initially it experiences no absorption.  All the population is initially in level $\ket{1}$ of Fig~\ref{4level}, which is a necessary requirement for STIRAP.  

The pump pulse is applied at a frequency outside this transmission window and the frequency difference between the Stokes and pump pulses matches the ground state splitting of the system, $\Delta_{g}$.  After STIRAP has modified the population distribution the absorption of a weak probe pulse provides a measurement of the population difference between the final level $\ket{3}$ and the optical excited state.  

The temporal profiles of the Stokes and pump pulses are nominally Gaussians, but there was a saturation effect from one of the AOMs with altered the shape of the Stokes pulse thus changing the Gaussian profile.  The FWHM of intensity for the two pulses was nominally 30/$\sqrt{2}$~$\mu$s.  The delay between the pump and Stokes pulses is variable, but is -17~$\mu$s, unless otherwise stated, with the Stokes pulse preceeding the pump pulse.  The probe pulse has a rectangular profile with a pulse duration of 1~ms, during which the frequency is scanned over 20~MHz, and is applied 300~$\mu$s after the pump pulse. In the interaction region of the crystal the pump, preparation and Stokes beams all have a beam diameter of 110~$\mu$m.  
 
\begin{figure}[!ht]
\centering
\includegraphics[width=9cm]{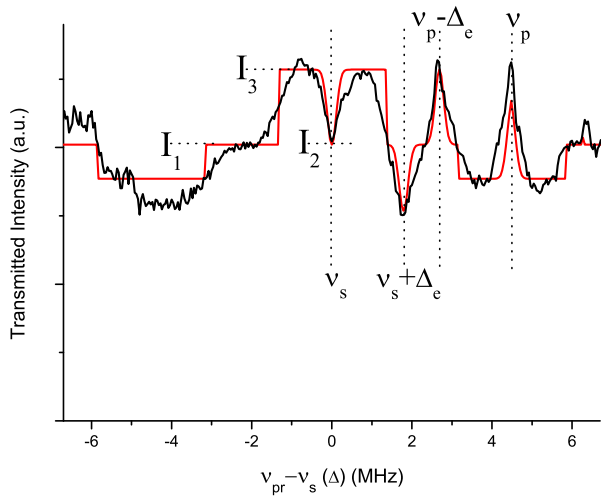}
\caption{(colour online) A typical experimental scan of the transferred population where $\nu_{pr}$ is the probe frequency, $\nu_{s}$ is the Stokes frequency, $\nu_{p}$ is the pump frequency and $\Delta_{e}$ is the excited state hyperfine splitting.  The ground state splitting was $\Delta_{g}$ = 4.55~MHz with an associated excited state splitting of $\Delta_{e} = $ 1.82~MHz.  The Rabi frequency of the strong transition was 1~MHz.  The black line is the experimental output whilst the red line is a fit to the experimental scan for this particular optical density.}\label{spec}
\end{figure}

Shown in Fig.~\ref{spec} is a typical experimental spectrum recording the transmission of the weak probe pulse as its frequency is scanned.  The overlapping red line is the expected probe transmission for this particular optical density after the preparation and STIRAP pulse sequences have been applied.  In this case the ground state splitting was $\Delta_{g}$ = 4.55~MHz and the excited state splitting was $\Delta_{e}$ = 1.82~MHz.

The various trenches of either increased or decreased transmission are related to the initial preparation step of the sequence.   After the preparation sequence there is increased transmission around the Stokes frequency, $\nu_{s}$, associated with the population having been optically pumped into the other ground state hyperfine level.  Correspondingly there is decreased transmission around the pump frequency, $\nu_{p}$, associated with the increased population in this level.  The decreased transmission also seen around $\nu_{s}-\Delta_{g}$ is due to the symmetrical nature of the holeburning spectrum.  

The preparation sequence actually creates two ensembles of thulium ions participating in the STIRAP sequence.  One ensemble is that displayed in Fig.~\ref{4level} where the Stokes pulse is applied on the strong transition.  The other ensemble is the opposite case, where the Stokes pulse is nearly resonant with the weak transition.  In each ensemble the situation is the opposite for the pump pulse.  

The STIRAP sequence of the temporally overlapping Stokes and pump pulses modifies this initial population distribution and the associated STIRAP absorption and transmission features are observed overlaid on the preparation trenches.  

The main STIRAP absorption feature is observed at the Stokes frequency which contains contributions from both ensembles of ions.  A STIRAP absorption feature is also seen at the Stokes frequency plus the excited state splitting, $\nu_{s}+\Delta_{e}$, which contains contributions from the ensemble where the Stokes pulse interacts with a strong transition.  There should also be a STIRAP absorption feature at $\nu_{s}-\Delta_{e}$, but the contributions to this feature only occur from the ensemble where the Stokes pulse interacts with a weak transition and thus is much smaller than the features at $\nu_{s}$ and $\nu_{s}+\Delta_{e}$.  

The corresponding STIRAP transmission features are observed at the pump frequency and at the pump frequency minus the excited state splitting, $\nu_{p}-\Delta_{e}$.  Again the absorption feature which should be observed at $\nu_{p}+\Delta_{e}$ contains contributions solely from the ensemble where the pump pulse interacts with a weak transition.

The important intensity levels are labelled I$_{1}$, I$_{2}$ and I$_{3}$ in Fig.~\ref{spec}.  I$_{1}$ corresponds to equal population in the two hyperfine ground state levels, $\ket{1}$ and $\ket{3}$, I$_{2}$ corresponds to the amount of transferred population in level $\ket{3}$ after the STIRAP sequence, whilst I$_{3}$ corresponds to zero population in the final level $\ket{3}$.  With these three intensity levels it is thus possible to determine the transfer efficiency using Beer's law.  In this case the transfer efficiency was $\sim$94\%.

\begin{figure}[!ht]
\begin{centering}
\includegraphics[width=6cm]{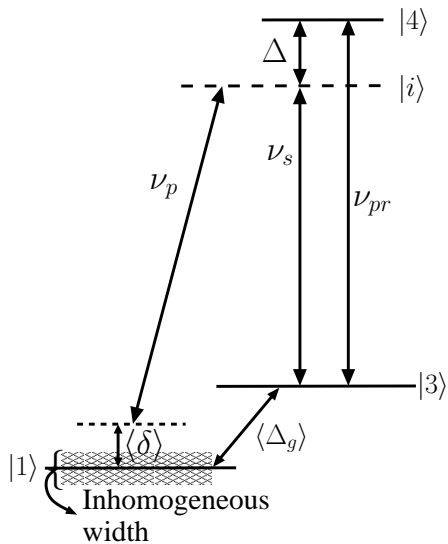}
\caption{Experimental level scheme, this is the simplified scheme of that shown in Fig.~\ref{4level} where we have removed level $\ket{2}$ for ease of reference.  The Stokes and pump pulses are optically resonant with a particular $\Lambda$ scheme comprising of levels $\ket{1}$, $\ket{3}$ and $\ket{i}$.  The probe pulse is then scanned and interacts with different $\Lambda$-systems with optical detunings of $\Delta$, determined during the STIRAP step.  The ground state splitting and two-photon detuning values, $\langle\Delta_{g}\rangle$ and $\langle\delta\rangle$, are average values due to the Raman inhomogeneous braodening present in our system.}\label{exp_levels}
\end{centering}
\end{figure}

The level scheme in Fig.~\ref{exp_levels} demonstrates how our method of optical detection functions.  The STIRAP experiment performed here is actually a two step process.  First we perform STIRAP.  The two-photon detuning and the ground state splitting are related to the pump-Stokes detuning by $\Delta_{g}-\delta = \nu_{p}-\nu_{s}$.  The ground state splitting is fixed by the applied magnetic field amplitude.  However, due to the inhomogeneous width of the ground state hyperfine transition in our system, the external field can only set an average value, $\langle\Delta_{g}\rangle$.  Therefore, the given pump and probe frequencies can only specify an average value of the two-photon detuning, given by $\langle\delta\rangle=\langle\Delta_{g}\rangle+\nu_{s}-\nu_{p}$. 

In the second step we scan the frequency of the weak probe pulse, $\nu_{pr}$.  The probe field precisely selects the ions with a definite value of the STIRAP optical detuning, $\Delta$, given by $\Delta$ = $\nu_{pr}-\nu_{s}$.  Indeed the probing spectral resolution is given by the homogeneous width, which amounts to a few kHz.  When $\nu_{pr}=\nu_{s}$ we are at optical resonance and the maximum transfer efficiency is observed.    

In Pr$^{3+}$:Y$_{2}$SiO$_{5}$ the inhomogeneous broadening of the hyperfine transition is small with respect to the practical Rabi frequencies.  Therefore, as far as STIRAP is concerned, the two-photon detuning is precisely defined by the pump and Stokes frequency difference, $\nu_{p}-\nu_{s}$.  Specifically, appropriate adjustment of this frequency difference can lead to two-photon resonance \cite{Klein2007a, Goto2006}.  In Tm$^{3+}$:YAG, the Raman inhomogeneous width grows with the magnitude of the applied magnetic field at the large rate of $\sim$400~kHz/T.  Hence the incidence of this inhomogeneous broadening on the STIRAP efficiency has to be examined. 

\section{Results and Discussion}

We will first present initial investigations of STIRAP in Tm$^{3+}$:YAG, characterising the transfer efficiency and the one- and two-photon detuning scans.  Following these initial STIRAP investigations we will present numerical and experimental data demonstrating the effect of the inhomogeneous width and Rabi frequency on both the maximum transfer efficiency and the two-photon linewidth of this system.

\begin{figure}[!ht]
\begin{centering}
\includegraphics[width=9cm]{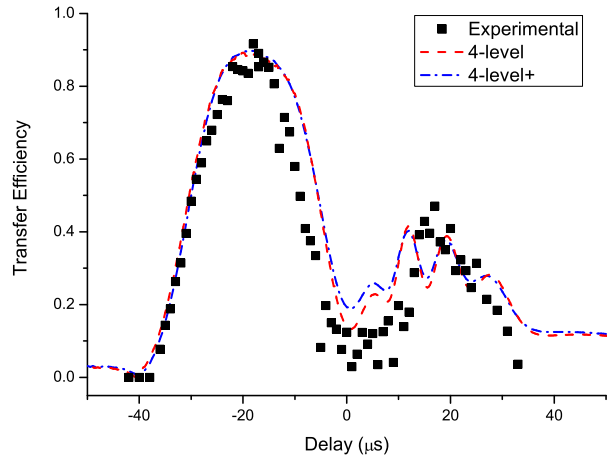}
\caption{(colour online) We vary the delay between the Stokes and pump pulses.  The dashed line (red) is the delay scan of the standard 4-level model whilst the dash-dot line (blue) is the 4-level model where we have included the inhomogeneous width to the Raman transition, henceforth to be labelled 4-level+, and the squares are the experimental data.  The ground state splitting was $\Delta_{g}$ = 4.55~MHz, the inhomogeneous width of the Raman transition was $\sim$50~kHz, the Rabi frequency of the strong transition was 1~MHz and the pulse widths were 30/$\sqrt{2}$~$\mu$s.  The simulations were performed at both optical and two-photon resonance.}\label{scan}
\end{centering}
\end{figure}

In our first experiment we varied the delay between the Stokes and pump pulses, covering both the Stokes preceding the pump, negative delays, and the pump preceeding the Stokes, positive delays.  The resulting transfer efficiencies were monitored for these different pulse orderings and are the squares displayed in Fig~\ref{scan}.  

We see that for optimum negative delays the experimental transfer efficiency reaches nearly 100\%.  Significant population transferral is also observed for positive delays, where the pump precedes the Stokes pulse.  In both these cases the population transferral is significantly larger than that transferred when the pump and Stokes pulses are applied coincident. The maxima in the experimentally measured transferred population show the characteristic plateaus which are a signature of the STIRAP process.  These plateaus imply that the transferral process is robust against small pulse variations.   

In Fig.~\ref{scan} we see that the maximum transfer efficiency occurs when the pulse separation of the Stokes and pump pulses is equal to the pulse width, at delay$\sim$-21~$\mu$s.  It has been previously discussed \cite{Gaubatz1990, Bergmann1998} that the maximum transfer efficiency is to be expected at this pulse separation for Gaussian shaped pulses as the adiabacity condition of Eqn.~\ref{adiabat}, taking $\Delta=0$ for simplicity, is best satisfied throughout the interaction at this pulse delay.  Thus we have the lowest losses via non-adiabatic coupling to the `leaky' dressed states $\ket{a^{\pm}}$ and the maximum transfer efficiency.   

The significant population transferral for positive delays is consistent with $b$-STIRAP \cite{Klein2007a, Vitanov1997, Vitanov1997b}.  In $b$-STIRAP the system remains in the bright state and the excited state is populated during the interaction.  Radiative losses are therefore possible resulting in a reduction in the transfer efficiency.  $b$-STIRAP is only observed in systems where the excited state radiative lifetime is long compared to the interaction time.  In Tm$^{3+}$:YAG the lifetime of the excited state is 800~$\mu$s \cite{Macfarlane1993}, which is much longer than the interaction time of the pulses.  

We see that the experimental data in Fig.~\ref{scan} agrees well with both the standard 4-level and the 4-level model where we have included the inhomogeneous broadening to the hyperfine transition, henceforth to be known as the 4-level+ model.  In performing these simulations and in subsequent modeling we have included the saturation effect of the AOM in our models.  

When including the significant inhomogeneous linewidth of the hyperfine transition in our 4-level+ model we assume the inhomogeneous line is a Gaussian centred on $\langle\delta\rangle$.  Our position in the inhomogeneous line is now specified as $\delta$ and thus for each value of $\langle\delta\rangle$ we must integrate over these $\delta$'s, whilst also including the Gaussian lineshape. The two-photon detuning profile in Fig.~\ref{Scan}, for the standard 4-level model, is the type of profile we use to perform this integration.  In order to obtain the transfer efficiency for the 4-level+ model at two-photon resonance, $\langle\delta\rangle = 0$, we apply a Gaussian filter to this profile, centred at $\delta=0$ and then integrate over the result.  By performing this integration process for a variety of delays we obtain the 4-level+ delay scan plotted here in Fig.~\ref{scan}.

In the situation in Fig.~\ref{scan} the Raman inhomogeneous linewidth is 50~kHz and the Rabi frequency on the strong transition is 1~MHz, with an associated weak Rabi frequency of 370~kHz.  Thus for these large Rabi frequencies and small inhomogeneous broadening we expect to be in the previously well-studied situation where the inhomogeneous broadening can be ignored \cite{Klein2007a, Goto2006}, as is demonstrated in these models.

The oscillations seen in the 4-level models for positive delays are thought to be due to the fact that the models were performed at $\Delta=0$, optical resonance.  At optical resonance it has been shown that the population transferral via $b$-STIRAP oscillates with the adiabacity parameter \cite{Vitanov1997b}, which in our case depends solely on the pulse delay between the pump and Stokes pulses.

\begin{figure}[!ht]
\begin{centering}
\includegraphics[width=9cm]{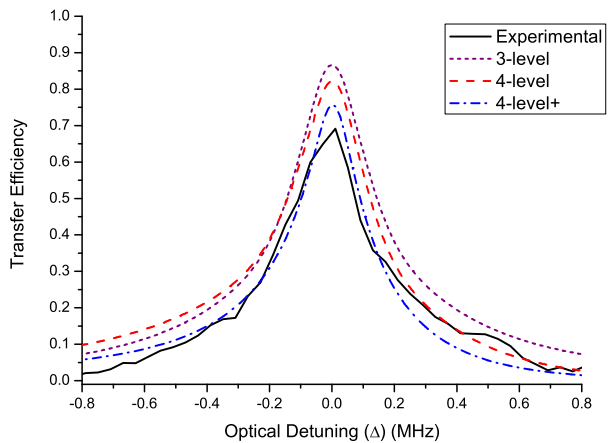}
\caption{(colour online) The solid (black) line shows the experimentally determined one-photon lineshape.  This lineshape is also determined with the standard 3-level model, dotted (purple) line, the standard 4-level model, dashed (red) line, and the 4-level model with the inhomogeneous linewidth of the hyperfine transition included, dash-dot (blue) line,  labelled as 4-level+.  The Rabi frequency of the strong transition was 510~kHz, the ground state splitting was 7.01~MHz and the inhomogeneous width of the Raman transition was 77.9~kHz.}\label{comparison}
\end{centering}
\end{figure}

In Fig.~\ref{comparison} is shown the one-photon linewidth for the various models along with that observed from experiment.  We see in this figure that the 4-level+ model agrees closely with the experimentally obtained result for this particular ground state splitting and Rabi frequency.

\begin{figure}[!ht]
\centering
\includegraphics[width=9cm]{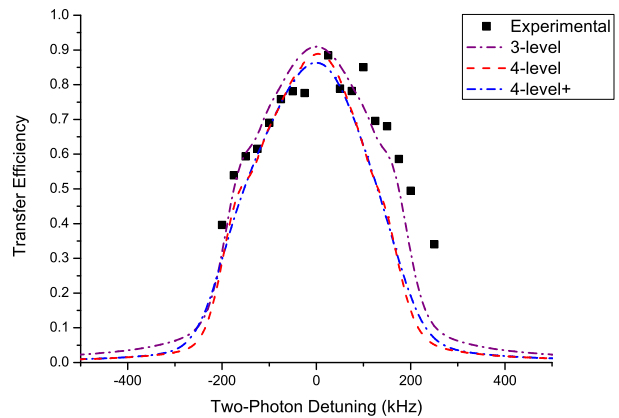}
\caption{(colour online) The two-photon detuning value is varied and the transfer efficiency is recorded experimentally, black squares.  The dotted (purple) line is the output from the standard 3-level model, the dashed (red) line is the output from the standard 4-level model and the dash-dot (blue) line is the output from the 4-level+ model, where in this case the two-photon detuning value plotted is actually the average two-photon detuning value.  The ground state hyperfine splitting was 7.1~MHz with an inhomogeneous width of 77.9~kHz and a Rabi frequency on the strong transition of 788~kHz. }\label{4levels}
\end{figure}

In Fig.~\ref{4levels} we compare the experimentally obtained two-photon detuning scan with that of the different models.  In the conditions of Fig.~\ref{4levels} the width of the two-photon detuning profile is much larger than the inhomogeneous broadening, of 77.9~kHz.  As a consequence,  the presence of the inhomogeneous line has little effect on the transfer efficiency.  The averaged profile given by the 4-level+ model hardly departs from the plain 4-level model.  The computed profiles closely agree with the experimental data.  In this experiment the strong Rabi frequency was 790~kHz, giving a Rabi frequency of 290~kHz on the weak transition, which is still much wider than the inhomogeneous width.

We now follow with an investigation into the effect the inhomogeneous width of the Raman transition has on the maximum transfer efficiency and the two-photon linewidth as we vary the applied Rabi frequency.  

\begin{figure}[!ht]
\begin{centering}
\includegraphics[width=8cm]{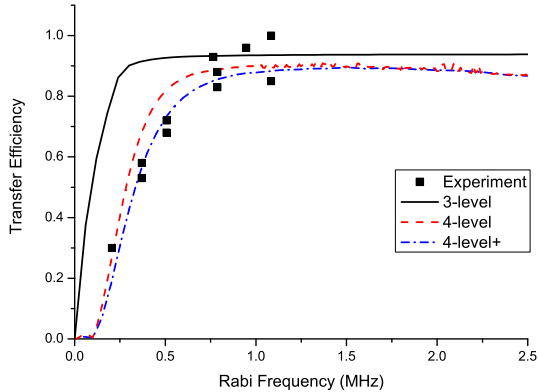}
\caption{(colour online) We vary the Rabi frequency whilst the inhomogeneous width and ground state splitting remain constant.  The splitting is constant at 7.1~MHz and the width of the inhomogeneously broadened Raman transition is 79~kHz.  The Rabi frequencies shown are those of the strong transition in each case.  The squares are the results obtained from experiments.  The solid (black) line is the simulation from the standard 3-level model, the dashed (red) line is the result from the standard 4-level model whilst the dash-dot (blue) line is the 4-level+ model.}\label{efficiency_Rabi}
\end{centering}
\end{figure}

Shown in Fig.~\ref{efficiency_Rabi} is the effect on the maximum transfer efficiency as the Rabi frequency is varied.  It is at low Rabi frequencies that the inhomogeneously broadened Raman transition is expected to have the largest effect on the transfer efficiency and we see that this is matched with the experimental and numerical results.

At low Rabi frequencies we expect a lower maximum transfer efficiency for the 4-level+ model when compared to that of the standard 4-level model.  In the standard 4-level model, at $\delta$=0, all the ions are at two-photon resonance.  Thus the majority of the population transferral occurs via efficient adiabatic transitions.  In the 4-level+ model, at $\langle\delta\rangle$=0, only a small portion of the ions are at two-photon resonance.  The rest will have a non-zero two-photon detuning value, dependant on their position in the inhomogeneous line.  Thus the majority of population transfer in this model occurs via the less efficient diabatic transitions, reducing the maximum transfer efficiency. As the Rabi frequency is increased a greater proportion of the transitions are adiabatic in the 4-level+ model, thus steadily increasing the maximum transfer efficiency, until it matches that of the standard 4-level model.

In this figure we also see a difference between the 3-level and 4-level models.  It is accepted that in a three-level system the population transfer is complete for a sufficiently high Rabi frequency satisfying the adiabacity criteria of either Eqns.~\ref{near_cond} or \ref{far_cond}.  The transfer process then simply becomes more robust as the Rabi frequency is increased.  This is seen in Fig.~\ref{efficiency_Rabi} where the transfer efficiency of the 3-level model rapidly reaches it's maximum value of $\sim$94\% and then remains constant as the Rabi frequency is increased.  The maximum transfer efficiency in this model is not 100\% due to the inclusion of relaxation and decoherence decay terms in the model.  

The situation is different for both the 4-level models where the transfer efficiency slowly reaches it's maximum value before then decreasing for larger Rabi frequencies.  

This reduction in the transfer efficiency might be due to the ``connectivity problem'', which has previously been observed in multi-level systems \cite{Shore1995,Martin1996,Martin1995}.  The connectivity problem means that whilst there may be an adiabtaic dressed state overlapping with the initial bare state at $t\rightarrow -\infty$ and another adiabatic dressed state overlapping with the desired final bare state at $t\rightarrow\infty$ there is no adiabatic or diabatic pathway connecting these two dressed states, due to the additional energy level and the multiple excitations of the transitions, resulting in no population transferral.  It has previously been demonstrated that for multi-level systems it is possible that the maximum transfer efficiency will reduce for large Rabi frequencies \cite{Martin1996, Shore1995} due to this problem.  It is also possible that at large Rabi frequencies the Stokes pulse could be strongly interacting with populated levels, resulting in optical pumping rather than STIRAP processes and thus reducing the transfer efficiency.   

Both experimental results and simulations show that the transfer efficiency, below that of maximum efficiency, is highly dependant on the Rabi frequency.  From the adiabacity condition of Eqn.~\ref{far_cond} we see that for a constant one-photon detuning and pulse duration that it is the variation of the Rabi frequency which will affect the transfer efficiency.  As the Rabi frequency is reduced a greater proportion of the population transferral will occur via non-adiabatic transitions, thus reducing the transfer efficiency due to the associated radiative losses. 

\begin{figure}[!ht]
\begin{centering}
\includegraphics[width=8cm]{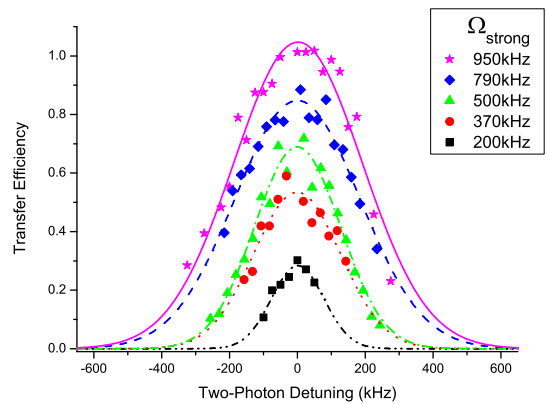}
\caption{(colour online) We varied the Rabi frequency for a constant ground state splitting whilst scanning the two-photon detuning.  The symbols are the experimentally obtained results whilst the lines are the fitted Gaussians.  The Rabi frequencies of the strong transition are listed for each scan.  The ground state splitting in each case was 7.3~MHz with an associated inhomogeneous width of the Raman transition of 80~kHz.}\label{Rabi_2photon}
\end{centering}
\end{figure}

We see explicitly in Fig.~\ref{Rabi_2photon} the dependence of the width of the two-photon detuning scan on the Rabi frequency.  We see that as we reduce the Rabi frequency, both the experimentally obtained maximum transfer efficiency and the two-photon detuning width is decreased.

Reducing the effective Rabi frequency will affect the maximum transfer efficiency in two ways. In order to achieve the maximum possible transfer efficiency the weak Rabi frequency should be much larger than the inhomogeneous width of the Raman transition and at the same time the adiabacity condition of Eqn.~\ref{far_cond} must be satisfied.  

Away from two-photon resonance population transferral is only possible via diabatic transitions.  As the Rabi frequency is reduced these diabatic transitions become weaker and hence it is thought that the weakening of these transitions reduces the two-photon width with decreasing Rabi frequency.   

\begin{figure}[!ht]
\begin{centering}
\includegraphics[width=8cm]{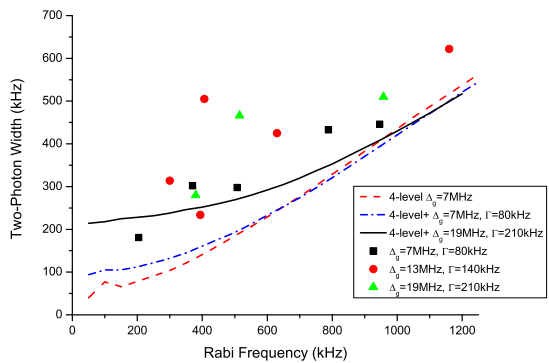}
\caption{(colour online) We varied the applied Rabi frequency and recorded the width of the two-photon detuning scan for different applied magnetic fields, and hence different Raman inhomogeneous widths.  The associated Raman inhomogeneous widths for the different ground state splitting values are $\Gamma\sim$80~kHz for $\Delta_{g}=$7~MHz, $\Gamma\sim$140~kHz for $\Delta_{g}=$13~MHz and $\Gamma\sim$210~kHz for $\Delta_{g}=$19~MHz, where $\Gamma$ is the width of the inhomogeneously broadened Raman transition.  The experimental two-photon detuning values, shown as the symbols, are the FWHM of a fitted Gaussian to the experimental scans.  The dashed (red) line is the standard 4-level model linewidths for a ground state splitting of $\Delta_{g}=$7~MHz.  The dash-dot (blue) and the solid (black) lines are the linewidths from the 4-level+ model for ground state splittings of $\Delta_{g}$=7~MHz and 19~MHz respectively.}\label{2photon_width}
\end{centering}
\end{figure}

We monitored in Fig.~\ref{2photon_width} the two-photon detuning width as we varied both the Rabi frequency and the inhomogeneous width of the system.  We found that the two-photon width was highly dependent on the Rabi frequency.  It was also dependent on the inhomogeneous width of the hyperfine transition, but less so and only at low Rabi frequencies.  

It was found numerically that there was a negligible difference between the two-photon linewidths from the standard 4-level models for different ground state splittings, thus only the result from a splitting of $\Delta_{g}$=7~MHz is displayed here.  Once the inhomogeneous width is included in the model we see that the two-photon width is dependent on the inhomogeneous width at low Rabi frequencies.  It is seen in this figure that the larger the inhomogeneous broadening and the smaller the Rabi frequency, the broader the two-photon linewidth of the 4-level+ model is when compared to that of the standard 4-level model.

The fact that the two-photon width is broader for the 4-level+ model over the standard 4-level model, at least at low Rabi frequencies is to be expected.  When the average two-photon detuning value is varied slightly in the 4-level+ model there can still be some ions at two-photon resonance, though less than if the average two-photon detuning was zero.  Whilst in the standard 4-level model, once there is a non-zero two-photon detuning value there are no ions at two-photon resonance.  Therefore there is greater population transfer in the 4-level+ model for non-zero two-photon detuning values, and hence a broader two-photon width.  This effect is larger for the model with larger Raman inhomogeneous broadening as is to be expected.  

As the Rabi frequency is increased the two-photon linewidths of the various models converge.  It is seen that the 4-level+ model with the smaller inhomogeneous width, $\Gamma$ = 80~kHz, converges at a smaller Rabi frequency than the model with a larger inhomogeneous width, $\Gamma$ = 210~kHz, where $\Gamma$ represnets the inhomogeneous linewidth.  We see that in each case the models with inhomogeneous broadening converge for a Rabi frequency of the strong transition of approximately six times the inhomogeneous width. 

This effect can also be seen in the experimental values where we see that for low Rabi frequencies the two-photon detuning widths appear to be dependent on the Raman inhomogeneous widths.  This effect appears to decrease as the Rabi frequency is increased.  

\section{Conclusions}

We have investigated STIRAP in Tm$^{3+}$:YAG both experimentally and numerically.  The experimental demonstrations presented here were the first such demonstrations of STIRAP in Tm$^{3+}$:YAG.  We observed robust population transfer with a maximum transfer efficiency of $\sim$90\% for STIRAP.  We were also able to observe $b$-STIRAP in this system due to the long population lifetime of the excited level as compared to the interaction time of the experiment.  Maximum transfer efficiencies of $\sim$45\% were observed for $b$-STIRAP, significantly larger than the transfer efficiency for coincident Stokes and pump pulses.

We were also able to completely model the Tm$^{3+}$:YAG system as this simple system consists of only 4 energy levels and a metastable state in a magnetic field.  We found that it was necessary to include the second excited state hyperfine level and the multiple excitations of the optical transitions in order for our numerical model to match the experimental results.  It was also seen that the inclusion of the Raman inhomogeneous broadening to our standard 4-level model altered the output and gave us simulated results more closely matched with those obtained experimentally, particularly at low Rabi frequencies.

We also discovered that the two-photon linewidth is broader when the inhomogeneous broadening is introduced.  This was thought to be due to more efficient population transferral at non-zero two-photon detuning values with the inhomogeneous broadening present.  It was seen that the two-photon linewidth determined from the model with inhomogeneous broadening converged with that of the standard 4-level model as the Rabi frequency was increased and the effect of the inhomogeneous broadening was reduced.  

It was also found that at low Rabi frequencies the presence of Raman inhomogeneous broadening reduced the maximum obtainable transfer efficiency.  But at high Rabi frequencies, ie. frequencies $\sim$10 times the inhomogeneous width, the inhomogeneous width made no difference to the obtained maximum transfer efficiency.

\end{document}